%% file: main.tex
\documentclass[10pt,twocolumn,letterpaper]{article}

\usepackage{cvpr}
\usepackage{times}
\usepackage{epsfig}
\usepackage{graphicx}
\usepackage{amsmath}
\usepackage{amssymb}
\usepackage{xcolor}
\usepackage{subcaption}
\usepackage{adjustbox}
\usepackage{algorithm}
\usepackage{algpseudocode}

\usepackage[breaklinks=true,bookmarks=false]{hyperref}

\DeclareMathOperator*{\argmin}{argmin} 
\DeclareMathOperator{\proj}{\textit{Proj}}

\cvprfinalcopy 

\def\cvprPaperID{10160} 

\ifcvprfinal\fi
\begin{document}

\title{GIFnets: Differentiable GIF Encoding Framework}

\author{Innfarn Yoo, Xiyang Luo, Yilin Wang, Feng Yang, and Peyman Milanfar\\ Google Research - Mountain View, California\\
\texttt{[innfarn, xyluo, yilin, fengyang, milanfar]@google.com}  
}

\maketitle
\thispagestyle{empty}

\begin{abstract}
Graphics Interchange Format (GIF) is a widely used image file format.
Due to the limited number of palette colors, GIF encoding often introduces color banding artifacts. Traditionally, dithering is applied to reduce color banding, but introducing dotted-pattern artifacts.
To reduce artifacts and provide a better and more efficient GIF encoding, we introduce a differentiable GIF encoding pipeline, which includes three novel neural networks: PaletteNet, DitherNet, and BandingNet.
Each of these three networks provides an important functionality within the GIF encoding pipeline. PaletteNet predicts a near-optimal color palette given an input image. DitherNet manipulates the input image to reduce color banding artifacts and provides an alternative to traditional dithering. Finally, BandingNet is designed to detect color banding, and provides a new perceptual loss specifically for GIF images.
As far as we know, this is the first fully differentiable GIF encoding pipeline based on deep neural networks and compatible with existing GIF decoders.
User study shows that our algorithm is better than Floyd-Steinberg based GIF encoding.
\end{abstract}

\input{1-introduction}
\input{2-previous_work}
\input{3-method}
\input{4-experiments}

\input{5-conclusion}

{\small
\bibliographystyle{ieee_fullname}
\bibliography{main}
}

\end{document}

%% file: 1-introduction.tex
\section{Introduction}

\begin{figure}
  \centering
  \begin{subfigure}{.49\linewidth}
    \centering
    \includegraphics[width=1.0\linewidth]{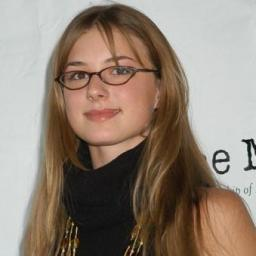}
    \caption{Original}
    \label{fig:teaser:a}
  \end{subfigure}
  \begin{subfigure}{.49\linewidth}
    \centering
    \includegraphics[width=1.0\linewidth]{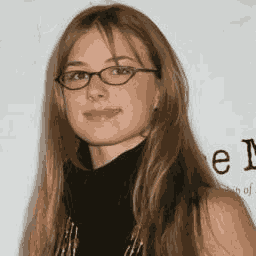}
    \caption{Palette}
    \label{fig:teaser:b}
  \end{subfigure}
  \\
  \begin{subfigure}{.49\linewidth}
    \centering
    \includegraphics[width=1.0\linewidth]{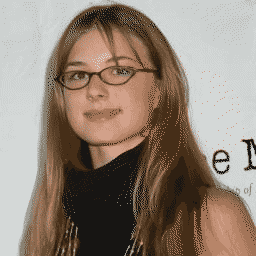}
    \caption{Floyd-Steinberg}
    \label{fig:teaser:c}
  \end{subfigure}
  \begin{subfigure}{.49\linewidth}
    \centering
    \includegraphics[width=1.0\linewidth]{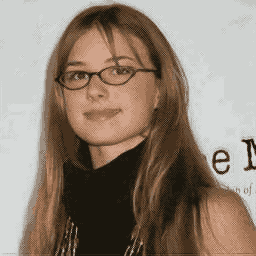}
    \caption{Our Method}
    \label{fig:teaser:d}
  \end{subfigure}
  \caption{Comparison of our method against GIF encoding with no dithering and Floyd-Steinberg dithering. Compared to GIF without dithering (b) and Floyd-Steinberg (c), our method (d) shows less banding artifacts as well as less dotted noise artifacts. The examples are generated with 16 palette colors.}
  \label{fig:teaser}
  \vspace{-0.1in}
\end{figure}

GIF is a widely used image file format. At its core, GIF represents an image by applying color quantization. Each pixel of an image is indexed by the nearest color in some color palette.
Finding an optimal color palette, which is equivalent to clustering, is an NP-hard problem. A commonly used algorithm for palette extraction is the median-cut algorithm~\cite{MedianCut}, due to its low cost and relatively high quality. Better clustering algorithms such as k-means produce palettes of higher quality, but are much slower, and have $\mathcal{O}(n^2)$ time complexity~\cite{celebi2011improving,pngquant}. Nearly all classical palette extraction methods involve an iterative procedure over all the image pixels, and are therefore inefficient.

Another issue with GIF encoding is the color banding artifacts brought by color quantization as shown in Figure~\ref{fig:teaser:b}. A popular method for suppressing banding artifacts is dithering, a technique which adds small perturbations in an input image to make the color quantized version of the input image look more visually appealing. Error diffusion is a classical method for dithering, which diffuses quantization errors to nearby pixels~\cite{Chang:2009:SED:1618452.1618508,floyd_steinberg,Ostromoukhov:2001:SEE:383259.383326}. While effective, these error diffusion methods also introduce artifacts of their own, \eg, dotted noise artifacts as shown in Figure~\ref{fig:teaser:c}. Fundamentally, these methods are not aware of the higher level semantics in an image, and are therefore incapable of optimizing for higher level perceptual metrics.



To this end, we propose a differentiable GIF encoding framework consisting of three novel neural networks: PaletteNet, DitherNet, and BandingNet.
Our architecture encompasses the entire GIF pipeline, where each component is replaced by one of the networks above. Our motivation for designing a fully differentiable architecture is two-fold. First of all, having a differentiable pipeline allows us to jointly optimize the entire model with respect to any loss functions. Instead of designing by hand heuristics for artifact removal, our network can automatically learn strategies to produce more visually appealing images by optimizing for perceptual metrics. Secondly, our design implies that both color quantization and dithering only involve a single forward pass of our network. This is inherently more efficient. Therefore, our method has $\mathcal{O}(n)$ time complexity, and is easy to be parallelized compared to the iterative procedures in the traditional GIF pipeline. 




\begin{figure*}[ht]
  \centering
  \begin{subfigure}{.95\linewidth}
    \centering
    \includegraphics[width=1.0\linewidth]{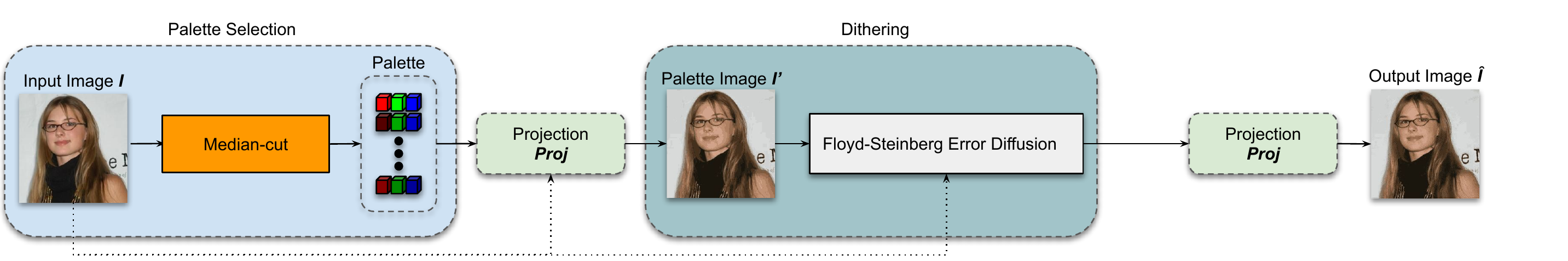}
    \caption{Traditional Pipeline}
    \label{fig:training_pipeline:a}
  \end{subfigure}
  \begin{subfigure}{.95\linewidth}
    \vspace{0.1 in}
    \centering
    \includegraphics[width=1.0\linewidth]{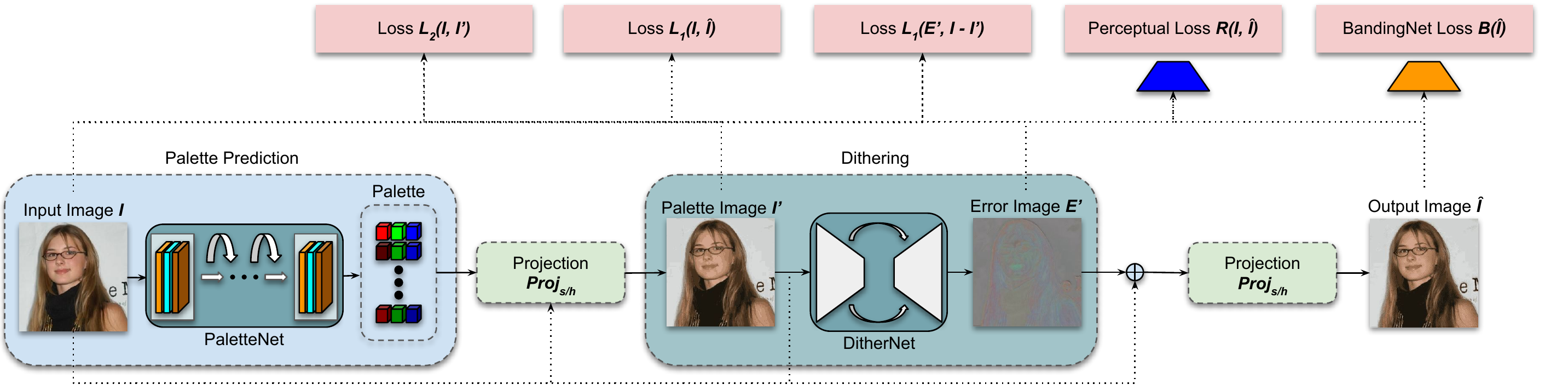}
    \caption{Our Pipeline}
    \label{fig:training_pipeline:b}
  \end{subfigure}
  \caption{Traditional GIF pipeline (a) and our GIF encoding pipeline (b). In our pipeline, PaletteNet predicts a near-optimal color palette and applies either a soft or hard projection to produce the quantized output image. DitherNet suppresses the banding artifacts by randomizing quantization errors, and BandingNet provides a perceptual loss to judge the severity of banding artifacts.}
  \label{fig:training_pipeline}
\end{figure*}

Our contributions are the following:

\begin{itemize}
  \item To the best of our knowledge, our method is the first fully differentiable GIF encoding pipeline. Also, our method is compatible with existing GIF decoders.
  \item We introduce PaletteNet, a novel method that extracts color palettes using a neural network. PaletteNet shows a higher peak signal-to-noise ratio (PSNR) than the traditional median-cut algorithm.
  \item We introduce DitherNet, a neural network-based approach to reducing quantization artifacts. DitherNet is designed to randomize errors more on areas that suffer from banding artifacts.
  \item We introduce BandingNet, a neural network for detecting banding artifacts which can also be used as a perceptual loss.
\end{itemize}

The rest of the paper is structured as follows: Section~\ref{sec:previous_work} reviews the current literature on relevant topics. Section~\ref{sec:method} gives a detailed explanation of our method. The experiments are given in Section~\ref{sec:experiments}, and conclusions in Section~\ref{sec:conclusion}.

%% file: 2-previous_work.tex
\section{Previous Work}
\label{sec:previous_work}

We categorize the related literature into four topics: dithering, palette selection, perceptual loss, and other related work.

\subsection{Dithering and Error Diffusion}
Dithering is a procedure to randomize quantization errors. Floyd-Steinberg error diffusion~\cite{floyd_steinberg} is a widely used dithering algorithm. The Floyd-Steinberg algorithm sequentially distributes errors to neighboring pixels.
Applying blue-noise characteristics on dithering algorithms showed improvement in perceptual quality~\cite{Ostromoukhov:2001:SEE:383259.383326,Ulichney1987}.
Kite \etal~\cite{841536} provided a model for error diffusion as linear gain with additive noise, and also suggested quantification of edge sharpening and noise shaping.
Adaptively changing threshold in error diffusion reduces artifact of quantization~\cite{892447}.

Halftoning or digital halftoning are other types of error diffusion methods, representing images as patternized gray pixels.
In halftoning, Pang \etal~\cite{Pang:2008:SH:1360612.1360688} used structural similarity measurement (SSIM) to improve halftoned image quality.
Chang \etal~\cite{Chang:2009:SED:1618452.1618508} reduced the computational cost of~\cite{Pang:2008:SH:1360612.1360688} by applying precomputed tables.
Li and Mould~\cite{doi:10.1111/j.1467-8659.2009.01596.x} alleviated contrast varying problems using contrast-aware masks.
Recently, Fung and Chan~\cite{6257485} suggested a method to optimize diffusion filters to provide blue-noise characteristics on multiscale halftoning, and Guo \etal~\cite{7165626} proposed a tone-replacement method to reduce banding and noise artifacts.

\subsection{Palette Selection}
Color quantization involves clustering the pixels of an image to $N$ clusters. One of the most commonly used algorithms for GIF color quantization is the median-cut algorithm~\cite{burger1989interactive}.
Dekker proposed using Kohonen neural networks for predicting cluster centers~\cite{dekker1994kohonen}. Other clustering techniques such as k-means~\cite{celebi2011improving}, hierarchical clustering~\cite{celebi2015effective}, particle swarm methods~\cite{omran2006particle} have also been applied to the problem of color quantization~\cite{scheunders1997comparison}.

Another line of work focuses on making clustering algorithms differentiable. Jang \etal~\cite{jang2016categorical} proposed efficient gradient estimators for the Gumbel-Softmax which can be smoothly annealed to a categorical distribution. Oord \etal~\cite{van2017neural} proposed VQ-VAE that generates discrete encodings in the latent space. Agustsson \etal~\cite{NIPS2017_6714} provides another differentiable quantization method through annealing. Peng \etal~\cite{peng2018k} proposed a differentiable k-means network which reformulates and relaxes the k-means objective in a differentiable manner.

\subsection{Perceptual Loss}
Compared to quantitative metrics, \eg, signal-to-noise ratio (SNR), PSNR, and SSIM, perceptual quality metrics measure human perceptual quality.
There have been efforts to evaluate perceptual quality using neural networks.
Johnson \etal~\cite{VggLoss} suggested using the feature layers of convolutional neural networks as a perceptual quality metric.
Talebi and Milanfar~\cite{NIMA} proposed NIMA, a neural network based perceptual image quality metric.
Blau and Michaeli~\cite{PercDistTradeoff} discussed the relationship between image distortion and perceptual quality, and provided comparisons of multiple perceptual qualities and distortions.
The authors extended their discussion further in~\cite{RethinkPercDistTradeoff} and considered compression rate, distortion, and perceptual quality.
Zhang \etal~\cite{Zhang_2018_CVPR} compared traditional image quality metrics such as SSIM or PSNR and deep network-based perceptual quality metrics, and discussed the effectiveness of perceptual metrics.

Banding is a common compression artifact caused by quantization.
There are some existing works~\cite{Baugh2014AdvancedDebanding,Lee2006TwoStageBanding,Wang2016PerceptualBanding} about banding artifact detection, where the method proposed by Wang \etal~\cite{Wang2016PerceptualBanding} achieved strong correlations with human subjective opinions.
Lee \etal~\cite{Lee2006TwoStageBanding} segmented the image into smooth and non-smooth regions, and computed various directional features, \eg, contrast and Sobel masks, and non-directional features, \eg, variance and entropy for each pixel in the non-smooth region to identify it as ``banding boundaries'' or ``true edges''. 
Baugh \etal~\cite{Baugh2014AdvancedDebanding} related the number of uniform segments to the visibility of banding. Their observation was that when the size of most of the uniform segments in an image was less than $10$ pixels in area, then there was no perceivable banding in the image.
Wang \etal~\cite{Wang2016PerceptualBanding} extracted banding edges from homogeneous segments, and defined a banding score based on length and visibility of banding edges. The proposed banding metrics have a good correlation with subjective assessment.

\subsection{Other Related Work}
GIF2Video~\cite{gif2video} is a recent work that also involves the GIF format and deep learning. GIF2Video tackles the problem of artifact removal for both static or animated GIF images. This is complementary to our work since GIF2Video is a post-processing method, whereas our method is a pre-processing method.

Our method also uses several well-known neural network architectures, such as ResNet~\cite{He_2016_CVPR}, U-Net~\cite{UNet}, and Inception~\cite{Szegedy_2015_CVPR}.
ResNet allowed deeper neural network connections using skip connections and showed significant performance improvement on image classification tasks. Inception used different convolution kernel sizes and merged outputs, and achieved better image classification accuracy than humans.
U-Net introduced an auto-encoder with skip connections, and achieved high quality output images in auto-encoder.

%% file: 3-method.tex
\section{Method}
\label{sec:method}

As shown in Figure~\ref{fig:training_pipeline:a}, GIF encoding involves several steps: (1) color palette selection; (2) pixel value quantization given the color palette; (3) dithering; (4) re-applying pixel quantization; (5) Lempel-Ziv-Welch lossless data compression. The last step is lossless compression, and it does not affect the image quality. Thus we will focus on replacing the first four steps with neural networks to improve the image quality. To make a differentiable GIF encoding pipeline, we introduce two neural networks: 1) PaletteNet, predicting the color palette from a given input image and 2) DitherNet for reducing quantization artifacts. We also introduce soft projection to make the quantization step differentiable. 
To suppress banding artifacts, we introduce BandingNet as an additional perceptual loss. Figure~\ref{fig:training_pipeline:b} shows the overall architecture of the differentiable GIF encoding pipeline.

\subsection{PaletteNet: Palette Prediction}
\label{ssec:method_palettenet}
The goal of PaletteNet is to predict a near-optimal palette with a feed-forward neural network given an RGB image and the number of palette $N_p$. We emphasize here that at inference time, the output from PaletteNet will be \emph{directly} used as the palette, and no additional iterative optimization is used. Therefore, PaletteNet in effect learns how to cluster image pixels in a feed-forward fashion.

We first state a few definitions and notations. Let $I\in\mathbb{R}^{H \times W \times 3}$ be the input image, $P\in \mathbb{R}^{N_p \times 3}$ a color palette with $N_p$ number of palettes, $Q(I): \mathbb{R}^{H \times W \times 3} \rightarrow \mathbb{R}^{N_p \times 3}$ be the palette prediction network. We define $I'$ to be the quantized version of $I$ using a given palette $P$, \textit{i.e.},
\begin{equation}
\label{eq:hard-palette-projection}
    I'[\alpha] = P[k], \quad k = \argmin_j |I'[\alpha] - P[j]|^2,
\end{equation}
where $\alpha$ is any indexing over the pixel space. 

Given the definitions above, PaletteNet is trained in a self-supervised manner with a standard $L_2$ loss.
\begin{equation}
\label{eq:palette-loss}
  \begin{split}
    L_{palette} = \frac{1}{H \times W} \sum_{\alpha}|I[\alpha] - P[k]|^2.
  \end{split}
\end{equation}
f
Next, we introduce the \emph{soft projection} operation to make the quantized image $I'$ also differentiable with respect to the network parameters of $Q$. Note that since the predicted palette $P$ is already differentiable with respect to $Q$, the non-differentiability of the projected image $I'$ is due to the hard projection operation defined in Equation~\ref{eq:hard-palette-projection}. Therefore, we define a soft projection $I'_s$ which smooths the $\argmin$ operation in Equation~\ref{eq:hard-palette-projection},
\begin{equation}
  \proj_s(I, P, \alpha) = I_{s}'[\alpha] = \sum_j w_j \cdot P[j],
\end{equation}
where $w_j = \frac{\exp(d_j / T)}{\sum_l \exp(d_l / T)}$, $d_j = \|P[j] - I_{s}'[\alpha]\|^2$, and $T$ a temperature parameter controlling the amount of smoothing. Note that the soft projection is not required for the training of PaletteNet itself, but needed if we want to chain the \emph{quantized} image $I'$ as a differentiable component of some additional learning system, such as in the case of our GIF encoding pipeline in Figure~\ref{fig:training_pipeline:b}.

\subsection{DitherNet: Reducing Artifacts}
\label{ssec:method_dithernet}

Dithering is an algorithm to randomize quantization errors into neighboring pixels in images to avoid banding artifacts.
After the color quantization, we define the error image as 
\begin{equation}
\label{eq:error_image}
E(\alpha) = I(\alpha) - \proj_h(I(\alpha), P),
\end{equation}
where $I$ is the original image, $\proj_h$ is a hard projection function that returns the nearest color in $P$, and $\alpha$ is any indexing over the pixel space. The traditional Floyd-Steinberg algorithm diffuses errors with fixed weights, and updates neighboring pixels in a sequential fashion.

Different from Floyd-Steinberg, our DitherNet $D$ accepts an original image $I$ and its quantized image $I'$ as an input, and directly predicts a randomized error image $E' = D(I, I')$. The predicted error is then added to the original image to produce the final quantized image.
\begin{equation}
\label{eq:I_hat}
\hat{I} = \proj_{s/h}(I + E', P),
\end{equation}
where $\proj_s$ is the soft projection and used for training to maintain differentiability. The hard projection ($\proj_h$) is used for inference.

For the network architecture, DitherNet uses a variation of U-Net~\cite{UNet} as a backbone network for creating error images. 
Training DitherNet is inherently a multi-loss problem. The final output has to remain close to the original, while achieving good visual quality after quantization. Ideally, DitherNet should be able to reduce the errors along the banding areas in a visually pleasing way. To this end, we define the dithering loss, $L_{dither}$, as a combination of image fidelity measures, perceptual losses, as well as a customized banding loss term to explicitly suppress banding artifacts. We define the $L_1$ loss as $L_1(x, y) = \sum|x(i) - y(i)|$, where $|\cdot|$ is the absolute difference operator and $i$ is the element index. The dither loss is defined as
\begin{equation}
  \label{eq:loss_dt}
  \begin{split}
    L_{dither} = \lambda L_1(I, \hat{I}) + \gamma L_1(E', I' - I) \\
    + \delta B(\hat{I}) + \theta R(I, \hat{I}),
  \end{split}
\end{equation}
where $B$ is the banding loss given by BandingNet (see Section~\ref{ssec:method_bandingnet}), $R$ is a perceptual loss given by either NIMA~\cite{NIMA} or VGG~\cite{VggLoss}, and $\lambda$, $\gamma$, $\delta$, and $\theta$ are weights of each loss.
In Equation~\ref{eq:loss_dt}, $L_1(I, \hat{I})$ preserves the similarity between input and final quantized images, $L_1(E', I - I')$ is for preserving the sum of quantization errors, and $B(\hat{I})$ is the banding loss.
We will discuss the effect of each term in Section~\ref{ssec:exp_dithernet}.

\subsection{BandingNet: Banding Score}
\label{ssec:method_bandingnet}

We propose a neural network that predicts the severity of banding artifacts in an image. We train the network by distilling from the non-differentiable banding metric in~\cite{Wang2016PerceptualBanding}. The output of this network will be used as the loss to guide our DitherNet.  

\begin{figure}[h]
  \centering
  \includegraphics[width=0.98\linewidth]{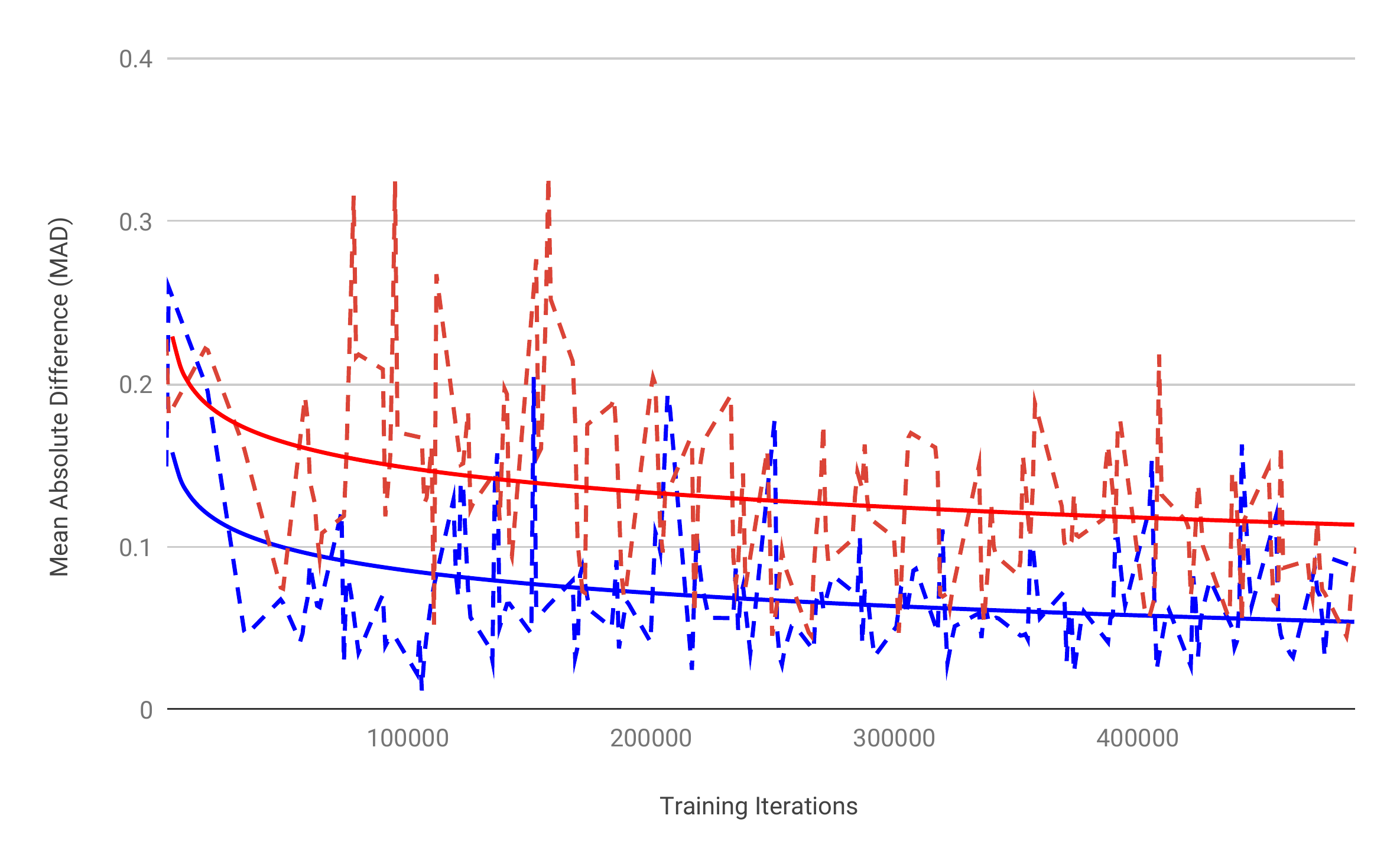}
  \caption{Mean Absolute Difference (MAD) between pre-computed and predicted banding scores. Red lines represent MAD without banding edge map, and blue lines show MAD with edge map over training iteration. Solid lines represent the trend of MAD over training iterations.}
  \label{fig:banding_mad}
\end{figure}

A straightforward way to train the model is to directly use the RGB image as input, and banding scores obtained by~\cite{Wang2016PerceptualBanding} as ground truth. We first tried to use a classical CNN model~\cite{NIPS2012_4824} to train the banding predictor, and defined the loss as the mean absolute difference (MAD) between predicted score and ground truth.
However, as shown in Figure~\ref{fig:banding_mad} (red lines), such naive approach is unstable over training, and could not achieve low MAD.

As pointed out in~\cite{Wang2016PerceptualBanding}, banding artifacts tend to appear on the boundary between two smooth regions. We found that adding a map of likely banding artifact areas as input would significantly improve the model. 
Here we propose a way to extract potential banding edges, and form a new input map (Algorithm~\ref{alg:banding_edgemap}).
\begin{algorithm}
\caption{Input Generation for BandingNet}\label{alg:banding_edgemap}
  \begin{algorithmic}[1]
  \Function{BandingInputs}{$RGB$}
    \State Converting $RGB$ to $YUV$
    \State Computing gradient $G_x$ and $G_y$ for $Y$ channel only
    \State Gradient map $G = ||{G_x}^2 + {G_y}^2||_2$
    \State Weight $W = ((1 - ReLU(1 - G))* ones(7\times{7}))^2$
    \State Banding edge map $E = W\cdot~G$
    \State \textbf{return} $M = [Y,~E]$
  \EndFunction
  \end{algorithmic}
\end{algorithm}

\begin{figure}[ht]
  \centering
  \begin{subfigure}{.325\linewidth}
    \centering
    \includegraphics[width=1.\linewidth]{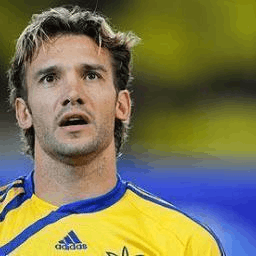}
    \caption{RGB}
  \end{subfigure}
  \begin{subfigure}{.325\linewidth}
    \centering
    \includegraphics[width=1.\linewidth]{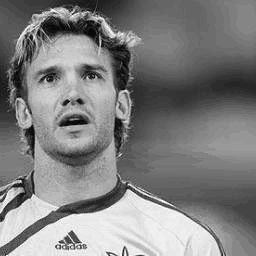}
    \caption{Y channel}
  \end{subfigure}
  \begin{subfigure}{.325\linewidth}
    \centering
    \includegraphics[width=1.\linewidth]{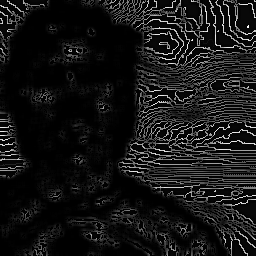}
    \caption{Extracted edge}
  \end{subfigure}
  \caption{Illustration of banding edge extraction.  The RGB image (a) is first converted to YUV, where the Y channel (b) is used to extract the banding edge map (c). Note that only edges in banding artifact areas are extracted.}
  \label{fig:banding_input_map}
\end{figure}

As shown in Figure~\ref{fig:banding_input_map}, the extracted edge map is able to highlight potential banding edges, \textit{e.g.}, banding on the background, and set relatively low weights for true edges like eyes and hair. By using this new input map, the training converges faster than using the RGB map, and MAD is also lowered as well, $0.057$ vs. $0.088$ within the banding score range [0, 1].

To use the BandingNet as a perceptual loss, we propose some additional modifications to the banding network proposed above. First, we augment the training data by adding pairs of GIF encoded images with and without Floyd-Steinberg dithering, and artificially lower the banding score loss for examples with Floyd-Steinberg dithering. This guides the loss along the Floyd-Steinberg direction, and can reduce adversarial artifacts compared to the un-adjusted BandingNet.
Secondly, we apply our BandingNet on a multi-scale image pyramid in order to capture the banding artifacts at multiple scales of the image. Compared to a single-scale loss, the multi-scale BandingNet promotes randomizing errors on a larger spatial range instead of only near the banding edges.
To define the multi-scale loss, we construct the image pyramid in Equation~\ref{eq:multi_scale_banding}.
\begin{equation}
    G(I, \eta) = F_{up}(F_{down}(S(I), \eta), 1/\eta),
    \label{eq:multi_scale_banding}
\end{equation}
where $G$ is a level of image pyramid, $F_{up}$ denotes image upscaling, $F_{down}$ denotes image downscaling, $S$ is a smoothing function, and $\eta$ is a scaling factor.  

\begin{figure*}[ht]
  \centering
  \begin{subfigure}{.245\linewidth}
    \centering
    \includegraphics[width=1.0\linewidth]{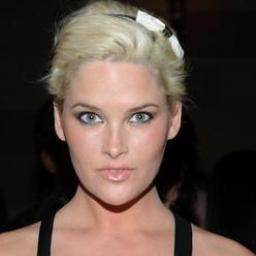}
  \end{subfigure}
  \begin{subfigure}{.245\linewidth}
    \centering
    \includegraphics[width=1.0\linewidth]{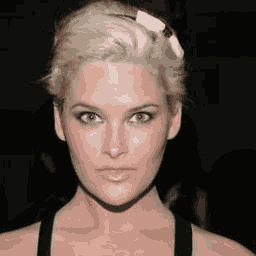}
  \end{subfigure}
  \begin{subfigure}{.245\linewidth}
    \centering
    \includegraphics[width=1.0\linewidth]{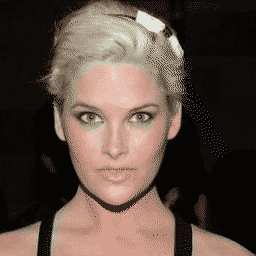}
  \end{subfigure}
  \begin{subfigure}{.245\linewidth}
    \centering
    \includegraphics[width=1.0\linewidth]{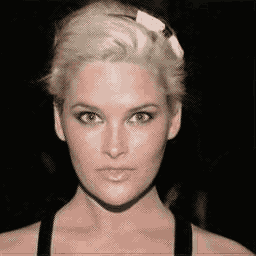}
  \end{subfigure}
  \\
  \vspace{0.025 in}
  \begin{subfigure}{.245\linewidth}
    \centering
    \includegraphics[width=1.0\linewidth]{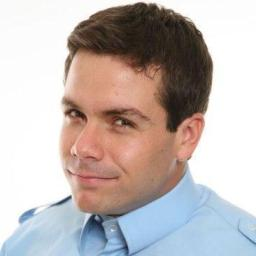}
  \end{subfigure}
  \begin{subfigure}{.245\linewidth}
    \centering
    \includegraphics[width=1.0\linewidth]{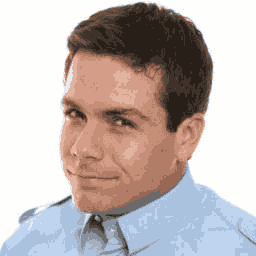}
  \end{subfigure}
  \begin{subfigure}{.245\linewidth}
    \centering
    \includegraphics[width=1.0\linewidth]{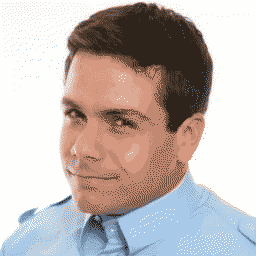}
  \end{subfigure}
  \begin{subfigure}{.245\linewidth}
    \centering
    \includegraphics[width=1.0\linewidth]{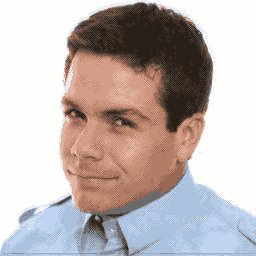}
  \end{subfigure}
  \\
  \vspace{0.025 in}
  \begin{subfigure}{.245\linewidth}
    \centering
    \includegraphics[width=1.0\linewidth]{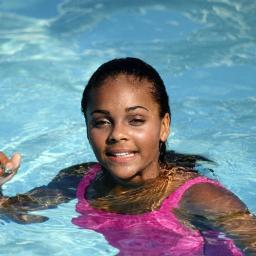}
    \caption{Original}
  \end{subfigure}
  \begin{subfigure}{.245\linewidth}
    \centering
    \includegraphics[width=1.0\linewidth]{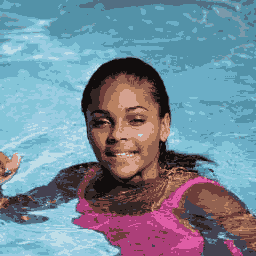}
    \caption{Palette}
  \end{subfigure}
  \begin{subfigure}{.245\linewidth}
    \centering
    \includegraphics[width=1.0\linewidth]{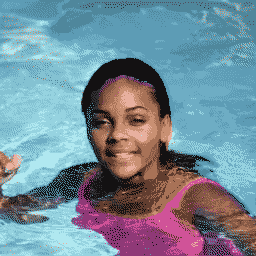}
    \caption{FS Dither}
  \end{subfigure}
  \begin{subfigure}{.245\linewidth}
    \centering
    \includegraphics[width=1.0\linewidth]{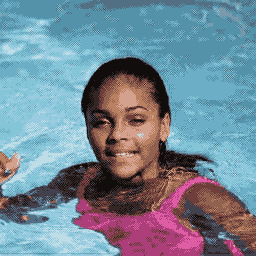}
    \caption{Ours}
  \end{subfigure}
  \caption{Sample images of our method compared to Median-cut and Floyd Steinberg. (a) original images, (b) quantized GIF with only PaletteNet, (c) Median-cut + Floyd-Steinberg, and (d) PaletteNet + DitherNet (our final method).}
  \label{fig:more_results_16}
\end{figure*}

Let $Z(I): \mathbb{R}^{H \times W \times 3} \rightarrow \mathbb{R}$ be the output of BandingNet. Our final banding loss is defined by
\begin{equation}
  \begin{split}
    B_{\eta}(I) = Z(G(I, \eta)) + Z(G(I, \eta^2)) \\
      + Z(G(I, \eta^3)) + Z(G(I, \eta^4))
    \label{eq:banding_loss}
  \end{split}
\end{equation}
For training DitherNet, we use $B_{1.5}(I)$. The exact training parameters for our BandingNet loss can be found in the supplementary materials.


\subsection{Overall Training}
\label{ssec:total_loss}

The overall loss for training our networks is given in Equation~\ref{eq:total_losses}.
\begin{equation}
\label{eq:total_losses}
\begin{split}
  L_{Total} = L_{palette} + L_{dither} = \\
    \beta L_2(I, I') + \lambda L_1(I, \hat{I}) + \gamma L_1(E', I - I') \\
    + \delta B(\hat{I}) + \theta R(I, \hat{I})
\end{split}
\end{equation}

To stabilize training, we propose a three-stage method for training all of the networks. In the first stage, BandingNet and PaletteNet are first trained separately until convergence. BandingNet will be fixed and used only as a loss in the next two stages. In the second stage, we fix PaletteNet and only train DitherNet using Equation~\ref{eq:total_losses}. In the final stage, we use a lower learning rate to fine-tune PaletteNet and DitherNet jointly. We found that the second stage is necessary to encourage DitherNet to form patterns that remove banding artifacts, whereas jointly training PaletteNet and DitherNet from the start will result in PaletteNet dominating the overall system, and a reduced number of colors in the output image.

%% file: 4-experiments.tex
\section{Experiments}
\label{sec:experiments}

We evaluate our methods on the CelebA dataset~\cite{liu2015faceattributes}, where the models are trained and evaluated on an 80/20 random split of the dataset. For both training and evaluation, we first resize the images while preserving the aspect ratio so that the minimum of the image width and height is $256$. The images are then center cropped to $256 \times 256$.

\subsection{PaletteNet}
\label{ssec:exp_palettenet}
For PaletteNet, we use InceptionV2~\cite{szegedy2016rethinking} as the backbone network, where the features from the last layer are globally pooled before passing to a fully connected layer with $N_p \times 3$ output dimensions. The output is then mapped to $[-1, 1]$ by the $\text{tanh}$ activation function.
To train PaletteNet, we use the $L_2$ loss in Equation~\ref{eq:palette-loss}. The details of training parameters are discussed in the supplementary material.

We evaluate the average PSNR of the quantized image compared to the original, for both median-cut and PaletteNet.
We also explored using PointNet~\cite{Qi_2017_CVPR}, a popular architecture for processing point cloud input, as a comparison.
The traditional method of palette extraction can be viewed as clustering a 3D point cloud, where each point corresponds to an RGB pixel in the image. The point cloud formed by the image pixels clearly retains rich geometric structures, and thus is highly applicable to PointNet. The details of PointNet are discussed in the supplementary material.
Results for different values of $N_p \in \{16, 32, 64, 128, 256\}$ are shown in Table~\ref{tab:psnr-palette-network}.
\begin{table}[h]
\centering
\resizebox{\columnwidth}{!}{%
\begin{tabular}{c |c c c c c} 
     PSNR           & $N_p=16$       & $N_p=32$      & $N_p=64$       & $N_p=128$      & $N_p=256$      \\ \hline
     Median-cut     & 28.10          & 30.78         & 33.27          & 35.61          & 37.80          \\ 
     PointNet       & 26.05          & 27.93         & 30.41          & 32.67          & 33.09          \\
     InceptionV2    & \textbf{29.24} & \textbf{31.6} & \textbf{33.75} & \textbf{35.81} & \textbf{38.08} \\ \hline 
\end{tabular}}
\caption{Average PSNR of quantized images for different palette extraction methods. Top row: Median-cut palette, Middle row: palette from PointNet, Bottom row: palette from InceptionV2.}
\label{tab:psnr-palette-network}
\end{table}

From Table~\ref{tab:psnr-palette-network}, we see that PaletteNet with the InceptionV2 network outperforms median-cut across all values of $N_p$, where the improvement is more pronounced for lower values of $N_p$. We also note that the PointNet architecture performed worse than the InceptionV2 network and Median-cut.

\subsection{DitherNet}
\label{ssec:exp_dithernet}


We note here that the DitherNet needs to be trained with relatively high weights on the various perceptual losses in order to produce perturbations that improve visual quality.
However, raising these weights too much introduces adversarial artifacts that lowers the perceptual loss, but no longer produces examples of high visual quality. To reduce this effect, we augment the input images by changing saturation, hue, and also apply early stopping where we terminate training at epoch 3.
The details of training parameters are discussed in the supplementary material.



\begin{figure}[h]
  \centering
  \begin{subfigure}{.32\linewidth}
    \centering
    \includegraphics[width=1.0\linewidth]{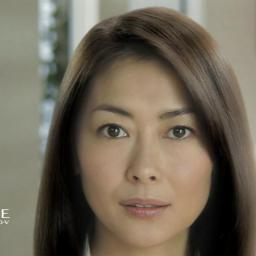}
    \caption{Original}
  \end{subfigure}
  \begin{subfigure}{.32\linewidth}
    \centering
    \includegraphics[width=1.0\linewidth]{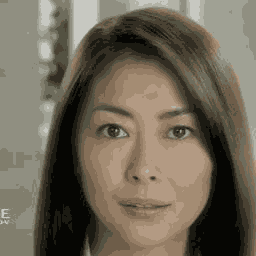}
    \caption{Without BL}
  \end{subfigure}
  \begin{subfigure}{.32\linewidth}
    \centering
    \includegraphics[width=1.0\linewidth]{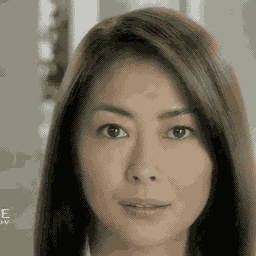}
    \caption{With BL}
  \end{subfigure}
  \caption{The original image before training (a), and the results of training without banding loss (b) and with banding loss (c).}
  \label{fig:with_and_without_banding_loss}
\end{figure}
\begin{figure}[ht]
  \centering
  \begin{subfigure}{.49\linewidth}
    \centering
     \includegraphics[width=1.0\linewidth]{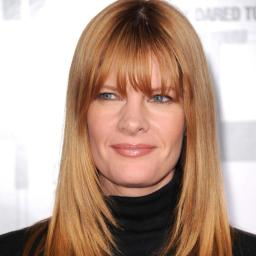}
    \caption{Original}
  \end{subfigure}
  \begin{subfigure}{.49\linewidth}
    \centering
    \includegraphics[width=1.0\linewidth]{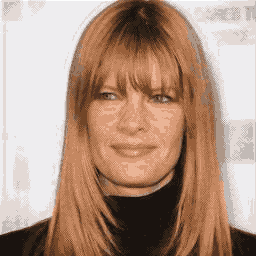}
    \caption{No Perc Loss}
  \end{subfigure}
  \\
  \begin{subfigure}{.49\linewidth}
    \centering
    \includegraphics[width=1.0\linewidth]{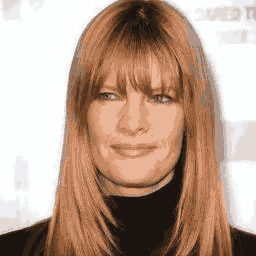}
    \caption{VGG Loss}
  \end{subfigure}
  \begin{subfigure}{.49\linewidth}
    \centering
    \includegraphics[width=1.0\linewidth]{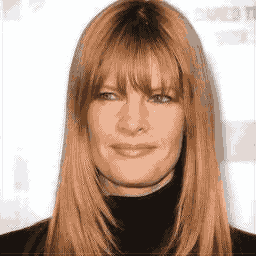}
    \caption{NIMA Loss}
  \end{subfigure}
  \caption{The results of training with and without perceptual loss ($N_p = 16$).}
  \label{fig:perceptual_loss}
\end{figure}

\begin{table}[h]
    \centering
    \begin{tabular}{c c c c}
                & Without Banding Loss  & With Banding Loss     \\ \hline
         PSNR   & 29.19                 & \textbf{28.65}        \\ \hline
         SSIM   & 0.854                 & \textbf{0.828}        \\ \hline 
    \end{tabular}
    \caption{PSNR and SSIM with and without banding loss for color palette with 16 colors.}
    \label{tab:banding_loss_effect}
\end{table}
Table~\ref{tab:banding_loss_effect} shows that training without banding loss provides better PSNR and SSIM.
However, DitherNet trained with banding loss provides better perceptual quality as shown in Figure~\ref{fig:with_and_without_banding_loss}.
In our experiment, our method shows better quality with VGG~\cite{VggLoss} and NIMA~\cite{NIMA} perceptual losses as shown in Figure~\ref{fig:perceptual_loss}.

\begin{figure*}[h]
  \centering
  \begin{subfigure}{.08\linewidth}
    \centering
    FS
  \end{subfigure}
  \begin{subfigure}{.17\linewidth}
    \centering
    \includegraphics[width=1.0\linewidth]{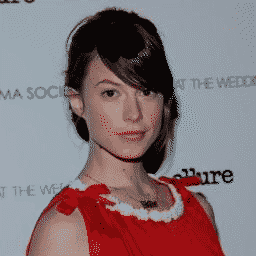}
  \end{subfigure}
  \begin{subfigure}{.17\linewidth}
    \centering
    \includegraphics[width=1.0\linewidth]{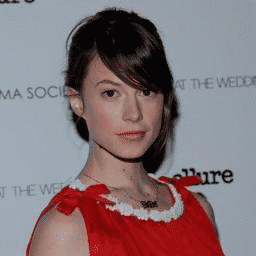}
  \end{subfigure}
  \begin{subfigure}{.17\linewidth}
    \centering
    \includegraphics[width=1.0\linewidth]{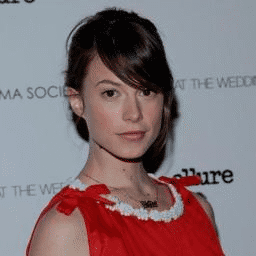}
  \end{subfigure}
  \begin{subfigure}{.17\linewidth}
    \centering
    \includegraphics[width=1.0\linewidth]{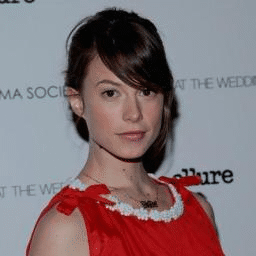}
  \end{subfigure}
  \begin{subfigure}{.17\linewidth}
    \centering
    \includegraphics[width=1.0\linewidth]{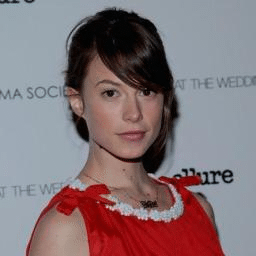}
  \end{subfigure}
  \\
  \begin{subfigure}{.08\linewidth}
    \centering
    Ours
  \end{subfigure}
  \begin{subfigure}{.17\linewidth}
    \centering
    \includegraphics[width=1.0\linewidth]{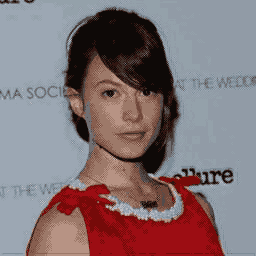}
  \end{subfigure}
  \begin{subfigure}{.17\linewidth}
    \centering
    \includegraphics[width=1.0\linewidth]{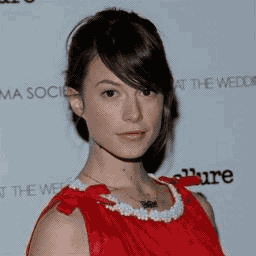}
  \end{subfigure}
  \begin{subfigure}{.17\linewidth}
    \centering
    \includegraphics[width=1.0\linewidth]{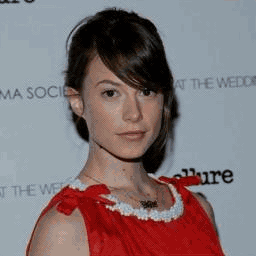}
  \end{subfigure}
  \begin{subfigure}{.17\linewidth}
    \centering
    \includegraphics[width=1.0\linewidth]{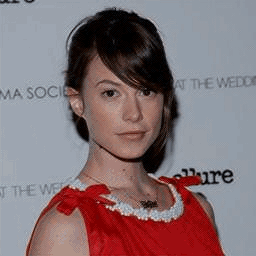}
  \end{subfigure}
  \begin{subfigure}{.17\linewidth}
    \centering
    \includegraphics[width=1.0\linewidth]{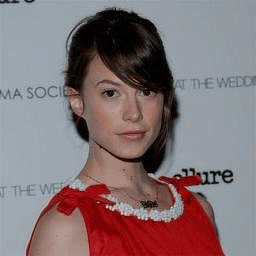}
  \end{subfigure}
  \\
  \vspace{0.05 in}
  \begin{subfigure}{.08\linewidth}
    \centering
    FS
  \end{subfigure}
  \begin{subfigure}{.17\linewidth}
    \centering
    \includegraphics[width=1.0\linewidth]{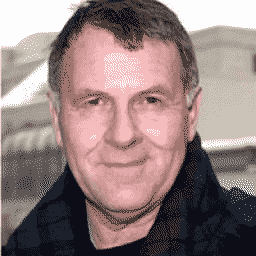}
  \end{subfigure}
  \begin{subfigure}{.17\linewidth}
    \centering
    \includegraphics[width=1.0\linewidth]{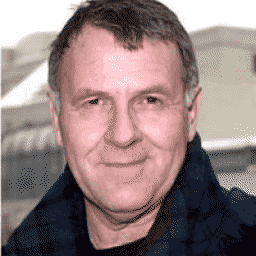}
  \end{subfigure}
  \begin{subfigure}{.17\linewidth}
    \centering
    \includegraphics[width=1.0\linewidth]{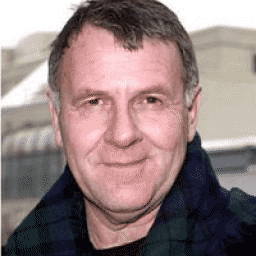}
  \end{subfigure}
  \begin{subfigure}{.17\linewidth}
    \centering
    \includegraphics[width=1.0\linewidth]{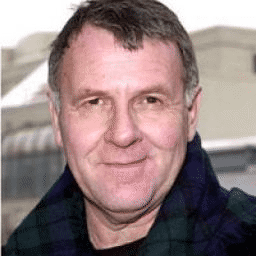}
  \end{subfigure}
  \begin{subfigure}{.17\linewidth}
    \centering
    \includegraphics[width=1.0\linewidth]{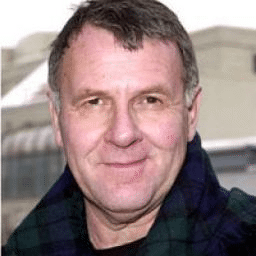}
  \end{subfigure}
  \\
  \begin{subfigure}{.08\linewidth}
    \centering
    Ours
  \end{subfigure}
  \begin{subfigure}{.17\linewidth}
    \centering
    \includegraphics[width=1.0\linewidth]{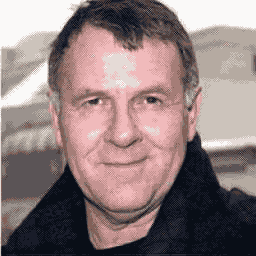}
    \caption{16 Palette}
  \end{subfigure}
  \begin{subfigure}{.17\linewidth}
    \centering
    \includegraphics[width=1.0\linewidth]{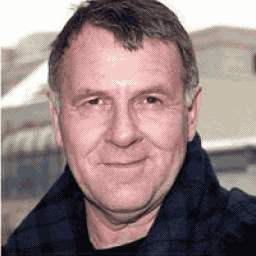}
    \caption{32 Palette}
  \end{subfigure}
  \begin{subfigure}{.17\linewidth}
    \centering
    \includegraphics[width=1.0\linewidth]{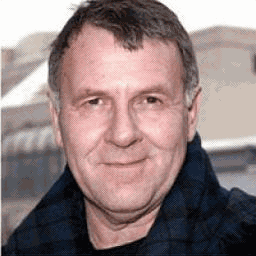}
    \caption{64 Palette}
  \end{subfigure}
  \begin{subfigure}{.17\linewidth}
    \centering
    \includegraphics[width=1.0\linewidth]{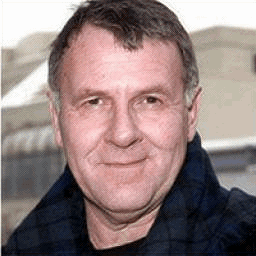}
    \caption{128 Palette}
  \end{subfigure}
  \begin{subfigure}{.17\linewidth}
    \centering
    \includegraphics[width=1.0\linewidth]{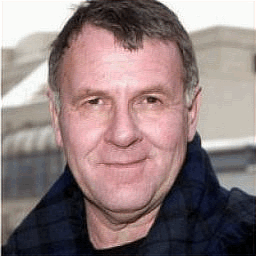}
    \caption{256 Palette}
  \end{subfigure}
  \caption{Side-by-side comparisons of our method and Floyd-Steinberg for 16, 32, 64, 128, and 256 palettes.}
  \label{fig:fs_vs_ours}
\end{figure*}

Quantitative metrics, such as PSNR or SSIM, are not proper metrics to compare dithering methods, since dithering is mainly used for improving perceptual quality and does not necessarily improve PSNR or SSIM over non-dithered images.
Instead, to evaluate the visual quality of our algorithm compared to the standard GIF pipeline with Floyd-Steinberg error diffusion and median-cut palette, a pairwise comparison user study was conducted using Amazon's Mechanical Turk. We randomly choose 200 images from the CelebA evaluation data, and produce quantized images from the standard GIF pipeline and our model. Raters are then asked to choose from a pair which image is of higher quality. For each image pair, we collect ratings from 10 independent users.

\begin{table}[h]
\centering
\begin{tabular}{c c c c c c} \\ \hline
  $N_p=16$       & $N_p=32$      & $N_p=64$  \\  
  87.5           & 85.0          & 79.4      \\ \hline 
  $N_p=128$      & $N_p=256$     & Average   \\
  45.1           & 47.1          & 68.8.     \\\hline
\end{tabular}
\caption{Favorability of our method compared to Median-cut + Floyd-Steinberg.}
\label{tab:user-study-results} 
\end{table}
Table~\ref{tab:user-study-results} shows the favorability of our method compared to the standard median-cut with Floyd-Steinberg dithering. We see that our method outperforms the baseline by a large margin when the number of palettes is low, and has comparable performance when the number of palettes is high. There are several causes to the discrepancy in favorability for different numbers of palette levels. First of all, the number of images with visible banding artifacts decreases as the number of palettes increases. On images without banding artifacts, our method is almost identical to that from the standard GIF pipeline, since raters are often not sensitive to the minute differences in PSNR.  On images with banding artifacts, a key difference between low and high palette count is in the visibility of the dotted pattern artifacts. When the number of palette levels is low, the dotted patterns are much more visible in the image and often rated unfavorably compared to the patterns from DitherNet. Another reason is the performance gap between PaletteNet and median-cut shrinks as the number of palettes grows (see Table~\ref{tab:psnr-palette-network}).

%% file: 5-conclusion.tex
\section{Conclusion}
\label{sec:conclusion}
In this paper, we proposed the first fully differentiable GIF encoding pipeline by introducing DitherNet and PaletteNet. To further improve the encoding quality, we introduced BandingNet that measures banding artifact score.
Our PaletteNet can predict high quality palettes from input images.
DitherNet is able to distribute errors and lower banding artifacts using BandingNet as a loss.
Our method can be extended in multiple directions as future work.
For example, k-means based palette prediction and heuristic methods for dithering, \textit{i.e.},~\cite{pngquant}, show higher visual quality than ours.
We also would like to extend our current work to image reconstruction, static to dynamic GIF, and connecting with other differentiable image file formats.